\documentclass[aps,prb,floatfix,twocolumn,showpacs,preprintnumbers,amsmath,amssymb,superscriptaddress]{revtex4-1}

\usepackage{graphicx}  
\usepackage{dcolumn}   
\usepackage{bm}        
\usepackage{stmaryrd}  
\usepackage{multirow}  
\usepackage{color}     
\usepackage{amsmath,amssymb}
\usepackage{enumerate}

\newcommand{\der}[3]{\frac{{\rm d}^{#1} #2}{{\rm d} #3^{#1}}}     
\newcommand{\av}[1]{ {\langle #1 \rangle} }                       
\newcommand{\Tr}[1]{\mathrm{Tr}\left\{ #1 \right\}}               

\def   \bhS      {{\hat{\bm S}}}                         
\def   \Ms       {M_{\rm s}}                             

\newcommand{\Heff}[1]{\mathbf{H}_{\mathrm{eff}\,#1}}     
\def   \happ     {H_{\rm app}}                           
\def   \hext     {H_{\rm ext}}                           
\def   \hani     {H_{\rm ani}}                           
\newcommand{\Hdem}[1]{\mathbf{H}_{\mathrm{dem}\, #1}}      
\newcommand{\Hint}[1]{\mathbf{H}_{\mathrm{int}\, #1}}      
\newcommand{\Ti}[1]{\bar{\mathbf{N}}_{#1}}               
\newcommand{\Hdx}[1]{H^{\rm d}_{#1 x}}                   
\newcommand{\Hdy}[1]{H^{\rm d}_{#1 y}}                   
\newcommand{\Hdz}[1]{H^{\rm d}_{#1 z}}                   
\newcommand{\Hix}[1]{H^{#1}_{x}}                         
\newcommand{\Hiy}[1]{H^{#1}_{y}}                         
\newcommand{\Hiz}[1]{H^{#1}_{z}}                         
\newcommand{\Hd}[1]{H^{\rm d}_{#1}}                      
\newcommand{\hrkky}[1]{H_{\mathrm{RKKY}\, #1}}           
\def   \Jrkky    {J_{\rm RKKY}}                          
\def   \Teff     {T_{\rm eff}}                           
\def   \Rsp      {R_{\rm sp}}                            
\def  \hsaf      {H_0}                                   
\def  \hsf       {H_0}                                   
\def  \Itr       {I_{\rm thr}}                            
\def  \PSD       {\mathrm{ PSD}}                         
\def  \RL        {R_{\rm L}}                             
\def  \bsig    {\mbox{\boldmath$\sigma$}}

\newcommand{\tor}[1]{{\bm\tau}_{#1}}                  
\newcommand{\torip}[1]{ {\bm \tau}_{#1 \perp} }       
\newcommand{\torop}[1]{ {\bm \tau}_{#1 \parallel} }   

\def   \du      {\downarrow\!\uparrow}
\def   \ud      {\uparrow\!\downarrow}
\newcommand{\Icdu}[1]{I^{\du}_{\mathrm{c}\, #1}}
\newcommand{\Icud}[1]{I^{\ud}_{\mathrm{c}\, #1}}

\def   \ex       {{\hat{\bm e}_{x}}}                     
\def   \ez       {{\hat{\bm e}_{z}}}                     
\newcommand{\ep}[1]{{\hat{\bm e}_{#1 \phi}}}             
\newcommand{\et}[1]{{\hat{\bm e}_{#1 \theta}}}           

\def   \gyro     {\gamma_{\rm g}}                        
\def   \eq       {{\rm e}}                               
\def   \kB       {k_{\mathrm{B}}}                        

\def   \nm       {{\rm nm}}                              
\def   \ns       {{\rm ns}}                              
\def   \Oe       {{\rm Oe}}                              
\def   \kOe      {{\rm kOe}}                             
\def   \K        {{\rm K}}                               
\def   \fOm      {{\rm f}\Omega{\rm m}^2}                
\def   \GHz      {\mathrm{GHz}}                          

\def  \AF        {{\mathrm{AF}}}                         
\newcommand{\N}[1]{\mathrm{N}_{#1}}                      
\newcommand{\F}[1]{\mathrm{F}_{#1}}                      

\begin{document}

\preprint{APS/123-QED}

\title{Current-induced dynamics of composite free layer\\ with antiferromagnetic interlayer exchange coupling}

\author{P.~Bal\'a\v{z}}
\affiliation{Department of Physics, Adam Mickiewicz University,
             Umultowska 85, 61-614~Pozna\'n, Poland}
\author{J.~Barna\'s}
\affiliation{Department of Physics, Adam Mickiewicz University,
             Umultowska 85, 61-614~Pozna\'n, Poland}
\affiliation{Institute of Molecular Physics, Polish Academy of Sciences
             Smoluchowskiego 17, 60-179 Pozna\'n, Poland}
\date{\today}

\begin{abstract}
  Current-induced dynamics in spin valves including composite free layer with antiferromagnetic
  interlayer exchange coupling is studied theoretically within the diffusive transport regime.
  We show that current-induced dynamics of a synthetic antiferromagnet is significantly different from dynamics of
  a synthetic ferrimagnet.  From macrospin simulations we obtain conditions for switching the composite free layer,
   as well as for appearance of
  various self-sustained dynamical modes.
  Numerical simulations are compared with simple analytical models of critical current
  based on linearized Landau-Lifshitz-Gilbert equation.
\end{abstract}

\maketitle


\section{Introduction}
\label{Sec:Introduction}

  After the effect of spin transfer torque (STT) in thin magnetic films
  had been predicted~\cite{Slonczewski1996:JMMM,Berger1996:PRB} and
  then experimentally proven~\cite{Tsoi1998:PRL,Katine2000:PRL},
  it was generally believed that current-controlled spin valve devices
  would replace soon the memory cells operated by external magnetic field.
  Such a technological progress, if realized, would certainly offer higher data
   storage density  and faster manipulation
  with the information stored on a hard drive memory.
  However, it became clear soon that some important issues must be solved before
  devices based on spin torque
  could  be used in practice.
  The most important is the reduction of current density needed for magnetic excitation (switching) in thin
  films,
  as well as enhancement of switching efficiency and thermal stability.
  Some progress has been made by using more complex spin valve structures
  and/or various subtle switching schemes based on optimized
  current and field pulses~\cite{Serrano-Guisan2008:PRL,Nikonov2010:JAP,Liu2009:APL,Balaz2009:PRB} .

  A significant enhancement of thermal stability can be achieved by replacing
  a simple free layer (single homogeneous layer)  with a system of two magnetic films separated by a thin nonmagnetic
  spacer,   known as composite free layer (CFL).
The spacer layer is usually thin so there is a  strong RKKY
exchange coupling between  magnetic
  layers~\cite{Grunberg1986:PRL,Parkin1991:PRB}.
  In practice, antiferromagnetic configuration is preferred
  as it reduces the overall magnetic moment of the CFL structure and
  makes the system less vulnerable to external magnetic fields
  and thermal agitation.
  When the antiferromagnetically coupled magnetic layers
  are identical, we call the structure synthetic antiferromagnet
  (SyAF). If they are different, then the CFL has uncompensated magnetic moment
  and such a system will be referred to as synthetic ferrimagnet (SyF).

  Current and/or field induced dynamics of CFLs is currently a
  subject of both experimental and theoretical investigations~\cite{JVKim2004:APL,Ochiai2005:APL,Smith2008:PRL,Lee2010:JMAG}.
  Switching scheme of SyAF  by magnetic field pulses has been proposed in Ref.~[\onlinecite{JVKim2004:APL}], and
  then the possibility of current-induced switching of SyAF
  was  demonstrated experimentally~\cite{Ochiai2005:APL}.
  In turn, the possibility of critical current reduction
   has been shown for a CFL with ferromagnetically coupled magnetic layers~\cite{Yen2008:APL}.
  However, the reduction of critical current in the case of antiferromagnetically coupled CFLs still  remains an open problem.
  In a recent numerical study on switching a SyAF free layer~\cite{You2010:JAP}
  it has been shown that the corresponding critical current in most cases is
  higher than the current required for  switching of a simple free layer, and only
  in a narrow range of relevant parameters (exchange coupling, layer thickness, etc.)
  the critical current  is reduced.
  Hence, proper understanding of current-induced dynamics of CFLs is highly required.
  We also note, that CFL can be used as a polarizer, too. Indeed, it has been shown
  recently~\cite{Mao2000:JAP}
  that SyAF used as a reference layer (with magnetic moment fixed
  fixed to adjacent antiferromangentic layer due to exchange anisotropy)
  might be excited due to dynamical coupling~\cite{Urazhdin2008:PRBR} with
  a simple sensing layer~\cite{Gusakova2009:PRB,Houssameddine2010:JAP}.

  The main objective of this paper is to study current-induced dynamics of a CFL
  with antiferomagnetic RKKY coupling in metallic spin valve pillars.
  We consider a system
  $\AF$/$\F{0}$/$\N{1}$/$\F{1}$/$\N{2}$/$\F{2}$, shown in Fig.~\ref{Fig:Scheme}, where
  $\AF$ is an antiferromagnetic layer (used to bias magnetization of the reference magnetic layer
  $\F{0}$),
  $\F{1}$ and $\F{2}$ are two magnetic layers,  while $\N{1}$ and $\N{2}$ are non-magnetic
  spacers. The part $\F{1}$/$\N{2}$/$\F{2}$ constitutes the CFL structure with antiferromagnetic
  interlayer exchange coupling.
  We examine current-induced dynamics of both SyAF and SyF free layers. These two structures differ only
  in the thickness of $\F{1}$ layer, while RKKY coupling and other pillar parameters
  remain the same.

  We assume that spin-dependent electron
  transport is diffusive in nature, and employ the model described in
  Refs.~[\onlinecite{Barnas2005:PRB}]
  and~[\onlinecite{GmitraBarnas:Chapt}].
   An important advantage of this model is the fact that it enables calculating
   spin current components and spin accumulation
   consistently in all magnetic and nonmagnetic
  layers, as well as current-induced torques exerted on all magnetic
  components.
  The torques acting at the internal interfaces of CFL introduce additional
  dynamical coupling between the corresponding magnetic layers.
  Consequently, the magnetic dynamics of CFL has been modelled by two
  coupled macrospins and described in terms of Landau-Lifshitz-Gilbert (LLG) equation.
  In addition, we derive some analytical expressions for critical
  currents from the stability conditions of linearized LLG equation in
  the static points~\cite{Bazaliy2004:PRB,Ebels2008:PRB}, and discuss results
  in the context of numerical simulations.

  The paper is organized as follows.
  In section 2 we describe the assumed  models for spin dynamics and STT calculations.
  In section~3 we analyze STT acting on CFL  and
  present results from numerical simulations on current-induced switching and
  magnetic dynamics. Some additional information on STT calculation can
  be found in the Appendix. Critical currents are derived and discussed in section~4.
  Finally, summary and general conclusions are in section~5.


\section{Magnetization dynamics}
\label{Sec:Dynamics}

  \begin{figure}[t]
    \centering
    \includegraphics[width=.9\columnwidth]{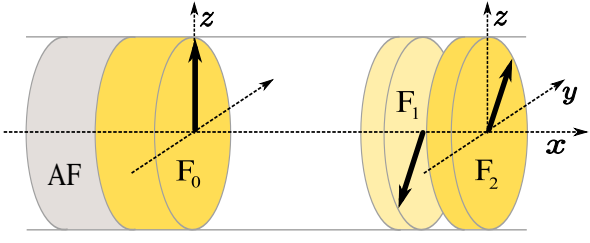}
    \caption{\label{Fig:Scheme}
             Scheme of the spin valve pillar structure with a composite free layer.}
  \end{figure}
In the macrospin approximation, magnetization dynamics of the CFL
  is described by two coupled Landau-Lifshitz-Gilbert (LLG)
  equations,
  \begin{equation}
  \label{Eq:LLG}
    \begin{split}
      &\qquad \der{}{\bhS_i}{t} + \alpha \bhS_i \times \der{}{\bhS_i}{t} = {\bm \Gamma}_i\,,\\
      &{\bm \Gamma}_i = -|\gyro| \mu_0 \bhS_i \times \Heff{i} + \frac{|\gyro|  }{\Ms d_i}\, \tor{i}\,,
    \end{split}
  \end{equation}
  for $i = 1, 2$, where $\bhS_i$ stands for a unit vector along the net spin moment of the $i$-th layer,
 whereas  $\Heff{i}$ and $\tor{i}$ are the effective field and current-induced torque, respectively,
 both acting on $\bhS_i$.
  The damping parameter $\alpha$ and the saturation magnetization $\Ms$  are assumed
  the same for both magnetic components
  of the CFL. Furthermore,
  $\gyro$ is the gyromagnetic ratio, $\mu_0$ is the vacuum permeability, and
  $d_i$ stands for thickness of the $\F{i}$ layer.

The effective magnetic field for the $\F{i}$ layer is
  \begin{equation}
  \label{Eq:Heff}
    \begin{split}
      \Heff{i} &= -\happ \ez - \hani (\bhS_{i} \cdot \ez) \ez + \Hdem{i}(\bhS_i) \\
               &\quad + \Hint{i}(\bhS_0, \bhS_j) + \hrkky{i}\, \bhS_j\,,
    \end{split}
  \end{equation}
  where $i,j = 1, 2$ and $i \neq j$.
  In the latter equation, $\happ$ is the external magnetic field applied  along the easy axis
  in the layers' plane (and oriented opposite to the axis $z$),
  $\hani$ is the uniaxial anisotropy field (the same for both magnetic
  layers), and
  $\Hdem{i} = (\Ti{i} \cdot \bhS_i) \Ms$
  is the self-demagnetization field of the $\F{i}$ layer with
  $\Ti{i}$ being the corresponding demagnetization tensor. Similarly,
  $\Hint{i} = (\Ti{0 i} \cdot \bhS_0) M_{\mathrm{s} 0} + (\Ti{j i} \cdot \bhS_j) \Ms$
  describes
  the  magnetostatic influence of the layers $\F{0}$ and $\F{j}$ on the layer $\F{i}$,
  respectively. Here, $M_{\mathrm{s} 0}$ is the saturation magnetization of the layer $\mathrm{F}_0$,
  which might be generally different from $\Ms$.
  Components of the tensors $\Ti{i}$, $\Ti{0 i}$ and $\Ti{i j}$ used in our simulations have been determined
by the numerical method introduced by Newell {\em et
al.\@}~\cite{Newell1993:JGR}.
  This method was originally developed for magnetic systems with non-uniform magnetization.
  To implement it into a macrospin model we
  considered discretized magnetic layers with uniform magnetizations,
  calculated tensors in each cell of the layer,
  and then averaged them along the whole layer.
  Since these tensors are diagonal, the demagnetization and magnetostatic fields can be expressed as
  $\Hdem{i} = (\Hdx{i} S_{i x}, \Hdy{i} S_{i y}, \Hdz{i} S_{i z})$, and
  $\Hint{i} = (\Hix{0 i} S_{0 x}, \Hiy{0 i} S_{0 y}, \Hiz{0 i} S_{0 z}) +
  (\Hix{j i} S_{j x}, \Hiy{j i} S_{j y}, \Hiz{j i} S_{j z})$, with
  $S_{i x}$, $S_{i y}$, and $S_{i z}$ denoting the components of
  the vector $\bhS_i$ ($i=0,1,2$) in the coordinate system shown in Fig.1.
  Finally, $\hrkky{i}$ stands for the  RKKY exchange field acting on $\bhS_i$,
  which is related to the RKKY coupling constant as
  $\hrkky{i} = -\Jrkky / (\mu_0 \Ms d_i)$~\cite{Gusakova2009:PRB}.

To include thermal effects we add
  to the effective field~(\ref{Eq:Heff}) a stochastic thermal field
  ${\mathbf H}_{{\rm th}\, i} = (H_{{\rm th}\, i x}, H_{{\rm th}\, i y}, H_{{\rm th\, i}
  z})$.
  For both spins its components obey the rules for Gaussian random
  processes
  $\av{H_{{\rm th}\, i \zeta}(t)} = 0$ and
  $\av{H_{{\rm th}\, i \zeta}(t) H_{{\rm th}\, j \xi}(t')} = 2 D \delta_{ij} \delta_{\zeta\xi} \delta(t - t')$,
  where $i,j=1,2$ and $\zeta,\xi=x,y,z$.
  Here, $D$ is the noise amplitude, which is related to the effective temperature, $\Teff$, as
  \begin{equation}
  \label{Eq:Dth}
    D = \frac{\alpha \kB \Teff}{\gyro \mu_0^2 \Ms V_i}\,,
  \end{equation}
  where $\kB$ is the Boltzmann constant, and $V_i$ is the volume of $\F{i}$ layer.

  In general, the current-induced torques acting on $\bhS_1$ and $\bhS_2$
  can be expressed as a sum of their in-plane and out-of-plane components
  $\tor{1} = \torop{1} + \torip{1}$ and $\tor{2} = \torop{2} + \torip{2} $, respectively.
  In a CFL structure, the layer $\F{1}$ is influenced by STT induced by  the polarizer $\F{0}$, as well as
  by STT due to the layer $\F{2}$.
  In turn, the layer $\F{2}$ is influenced by the torques from the layer $\F{1}$.
  Hence we can write
  \begin{subequations}
  \label{Eqs:STT}
    \begin{align}
      \torop{1} &= I\bhS_1 \times \left[ \bhS_1 \times \left( a_1^{(0)} \bhS_0 + a_1^{(2)} \bhS_2 \right) \right]\,, \\
      \torip{1} &= I\bhS_1 \times \left( b_1^{(0)} \bhS_0 + b_1^{(2)} \bhS_2 \right)\, , \\
      \torop{2} &= Ia_2^{(1)}\, \bhS_2 \times \left( \bhS_2 \times \bhS_1 \right)\, , \\
      \torip{2} &= Ib_2^{(1)}\, \bhS_2 \times \bhS_1\, ,
    \end{align}
  \end{subequations}
  where   $I$ is the charge current density, which is positive when
  electrons flow from the layer $\F{2}$ towards $\F{0}$  (see
  Fig.~\ref{Fig:Scheme}), while
  the parameters $a_{i}^{(j)}$ and $b_{i}^{(j)}$ ($i,j =1,2$) are independent of current $I$, but generally depend on
  magnetic configuration.

  We write the current density in the spin space as
${\bf j}=j_0{\bf 1}+{\bf j}\cdot \bsig$, where $j_0$ is the
particle current density ($I=ej_0$), ${\bf j}$ is the spin current
density (in the units of $\hbar /2$), $ \bsig$ is the vector of
Pauli matrices, and ${\bf 1}$ is a $2\times 2$ unit matrix.
  In frame of the diffusive transport model~\cite{Barnas2005:PRB},
  the parameters $a_{i}^{(j)}$ and $b_{i}^{(j)}$  are given by the formulas
  \begin{align}
    a_1^{(0)} &= -\frac{\hbar}{2}\, \frac{j'_{1y}\mid_{\rm \N{1}/\F{1}}}{I\sin\theta_{01}}\, , &
    b_1^{(0)} &= \frac{\hbar}{2}\, \frac{j'_{1x}\mid_{\rm \N{1}/\F{1}}}{I\sin\theta_{01}}\, , \notag \\
    a_1^{(2)} &= -\frac{\hbar}{2}\, \frac{j''_{2y}\mid_{\rm \F{1}/\N{2}}}{I\sin\theta_{12}}\, , &
    b_1^{(2)} &= \frac{\hbar}{2}\, \frac{j''_{2x}\mid_{\rm \F{1}/\N{2}}}{I\sin\theta_{12}}\, , \\
    a_2^{(1)} &= -\frac{\hbar}{2}\, \frac{j'''_{2y}\mid_{\rm \N{2}/\F{2}}}{I\sin\theta_{12}} \, , &
    b_2^{(1)} &= \frac{\hbar}{2}\, \frac{j'''_{2x}\mid_{\rm \N{2}/\F{2}}}{I\sin\theta_{12}}\, , \notag
  \end{align}
  where the angles $\theta_{01}$ and $\theta_{12}$ are given by
  $\cos\theta_{01} = \bhS_0 \cdot \bhS_1$ and $\cos\theta_{12} = \bhS_1 \cdot \bhS_2$.
  Here, $j'_{1y}$ and $j'_{1x}$ are transversal (to $\bhS_1$) components of spin current in
  the layer $\N{1}$ (taken at the $\N{1}$/$\F{1}$ interface) written in the local coordinate system of $\bhS_1$. Thus,
  $j'_{1y}$ is parallel to the vector $\bhS_1 \times (\bhS_1 \times \bhS_0)$
  (lying in the plane defined by $\bhS_0$ and $\bhS_1$) and
  $j'_{1x}$ is aligned with $\bhS_1 \times \bhS_0$
  (perpendicular to the plane defined by $\bhS_0$ and $\bhS_1$).
  Note, that the $z$-component of spin current is aligned along $\bhS_1$
  and does not contribute to STT~\cite{Stiles2002:PRBa}.
  Similarly, we define the  torque amplitudes acting inside the CFL.
  Here, $j''_{2y}$ and $j''_{2x}$ are the components of spin current in the  layer $\N{2}$
  (taken at the N{2}/F{1} interface), transversal to $\bhS_1$  and  lying respectively
  in the plane and perpendicularly to the plane
  defined by $\bhS_1$ and $\bhS_2$.
  Analogically, $j'''_{2y}$ and $j'''_{2x}$ are spin current components in $\N{2}$
  transversal to $\bhS_2$.
  The relevant  spin current transformations are given in Appendix~\ref{Sec:Transformations}.
  Note, that the spin current components depend linearly on $I$, so the parameters $a_{i}^{(j)}$ and
  $b_{i}^{(j)}$ are independent of the current density $I$.


\section{Numerical simulations}
\label{Sec:Simulations}

  In this section we present results on our numerical simulations of
  current-induced dynamics for two metallic pillar structures
  including CFL with antiferromagnetic interlayer exchange coupling.
  As described in the introduction, the considered pillars have the general structure
  $\AF$/$\F{0}$/$\N{1}$/$\F{1}$/$\N{2}$/$\F{2}$ (see Fig.(1)).
  More specifically, we consider spin valves
  Cu - IrMn(10)/Py(8)/Cu(8)/Co($d_1$)/Ru(1)/Co($d_2$) - Cu, where
  the numbers in brackets stand for the layer thicknesses in nanometers.
  The layer Py(8) is the Permalloy polarizing layer with its magnetization fixed due to
  exchange coupling to IrMn.
In turn, Co($d_1$)/Ru(1)/Co($d_2$) is the CFL
($\F{1}$/$\N{2}$/$\F{2}$ structure) with antiferromagnetic RKKY
exchange coupling {\it via} the thin ruthenium layer.
  The coupling constant between Co layers has been assumed as
  $\Jrkky \simeq -0.6\,\mathrm{mJ}/\mathrm{m}^2$,
  which is close to experimentally observed values~\cite{Gusakova2009:PRB,Houssameddine2010:JAP}.
  Here, we shall analyze two different geometries of CFL.
  The first one is a SyAF structure with $d_1 = d_2 = 2\,\nm$,
  while the second one is a synthetic ferrimagnet (SyF) with $d_1 = 2 d_2 = 4\,\nm$.

  Simulations have been based on numerical integration of the two coupled LLG equations~(\ref{Eq:LLG})
  with simultaneous calculations of STT, see Eq.(\ref{Eqs:STT}).
  We have assumed typical values of the relevant parameters, i.e.,
the damping parameter has been set to $\alpha = 0.01$, while the
uniaxial anisotropy field
  $\hani = 45\,\mathrm{kAm}^{-1}$ in both magnetic layers of the  CFL.
  In turn, saturation magnetization of cobalt has been assumed as
  $\Ms(\mathrm{Co}) = 1.42 \times 10^6\, \mathrm{Am}^{-1}$, and
  for permalloy $\Ms(\mathrm{Py}) = 6.92 \times 10^5\, \mathrm{Am}^{-1}$.
  The demagnetization field and magnetostatic interaction
  of magnetic layers have been calculated for layers of elliptical
cross-section, with the major and minor axes equal to $130\,\nm$
and $60\,\nm$, respectively.

  For both  structures under consideration we have analyzed the current-induced dynamics
  as a function of current density and external magnetic field.  The results have been presented
  in the form of diagrams displaying time-averaged values of the pillar resistance.
  Numerical integration of~Eq.(\ref{Eq:LLG}) has been
  performed using corrector--predictor Heun scheme, and
  the results have been verified for integration steps in the range from
  $10^{-4}\,\ns$ up to $10^{-6}\,\ns$.
  The STT components acting on CFL spins have been calculated at each integration step
  from the spin currents, which have been numerically calculated
  from the appropriate boundary conditions~\cite{Barnas2005:PRB}.
  Similarly, resistance of the studied pillars has been calculated from spin accumulation in
  frame of the model used also for the STT description
  (for details see also Ref.~\onlinecite{Gmitra2009:PRB}).

\subsection{Spin transfer torque}

  Let us analyze first the angular dependence of STT components in the structures under consideration.
  Although the thicknesses of
  magnetic layers in the studied SyAF and SyF structures are different,
  the angular dependence of STT components as well as their
  amplitudes are very similar. Thus, the analysis of STT in SyAF applies also qualitatively
   to the studied SyF free layer.

  \begin{figure}[h]
    \centering
    \includegraphics[width=8cm]{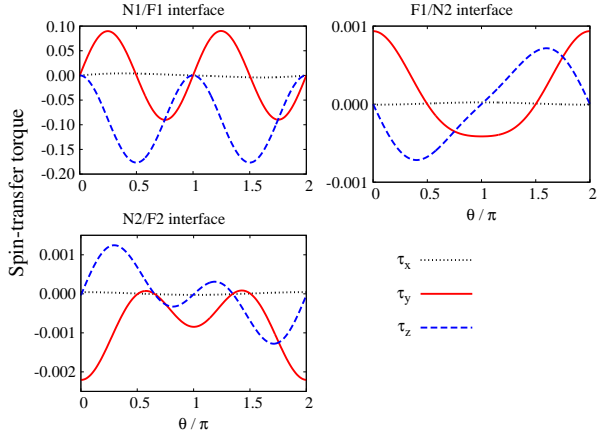}
    \caption{\label{Fig:stt1} (color online)
             Angular dependence of the cartesian components of STT, in the units of $\hbar I /
             |\eq|$,
             acting at N/F interfaces, when magnetization of the
             SyAF structure is rotated rigidly with both $\bhS_1$ and $\bhS_2$ remaining in the corresponding
             layer planes.
             Here, $\theta$ is an angle between $\bhS_1$ and $\ez$.
             $\bhS_2$ is tilted away from the antiparallel configuration with $\bhS_1$ by an
             angle of $1^{\circ}$.}
  \end{figure}

  First, we analyze STT components in the case when SyAF is rotated as a rigid structure,
  i.e. the antiparallel configuration of $\bhS_1$ and $\bhS_2$ is maintained.
  To have a nonzero torque between $\F{1}$ and $\F{2}$ layers, $\bhS_2$ has been tilted away from the
  antiparallel configuration by an angle of $1^{\circ}$.
  Figure~\ref{Fig:stt1} shows all three cartesian components (see Fig.1 for definition of the coordinate system)
  of STT acting at N/F interfaces as a
  function of the angle $\theta$  between $\ez$ and $\bhS_1$.
  While the $y$ and $z$-components are in the plane of the layers (the spins of CFL are rotated in the layer plane),
  the component $x$ is normal to the layer plane.
  However, $\tau_x$ remains negligible at all interfaces of the CFL.
  The STT acting at $\N{1}$/$\F{1}$ reveals a standard (non-wavy~\cite{Gmitra2006:PRL}) angular dependence,
  and vanishes when $\bhS_1$  is collinear with $\bhS_0$.
  Its amplitude is comparable to STT in standard spin valves with a simple free layer.
  The STT at $\F{1}$/$\N{2}$ and $\N{2}$/$\F{2}$ interfaces also depends on the angle $\theta$.
  However, they are about two order of magnitude smaller, which is
a consequence of a small angle ($1^{\circ}$) assumed between
$\bhS_1$ and $\bhS_2$.

\begin{figure}[t]
    \centering
    \includegraphics[width=8cm]{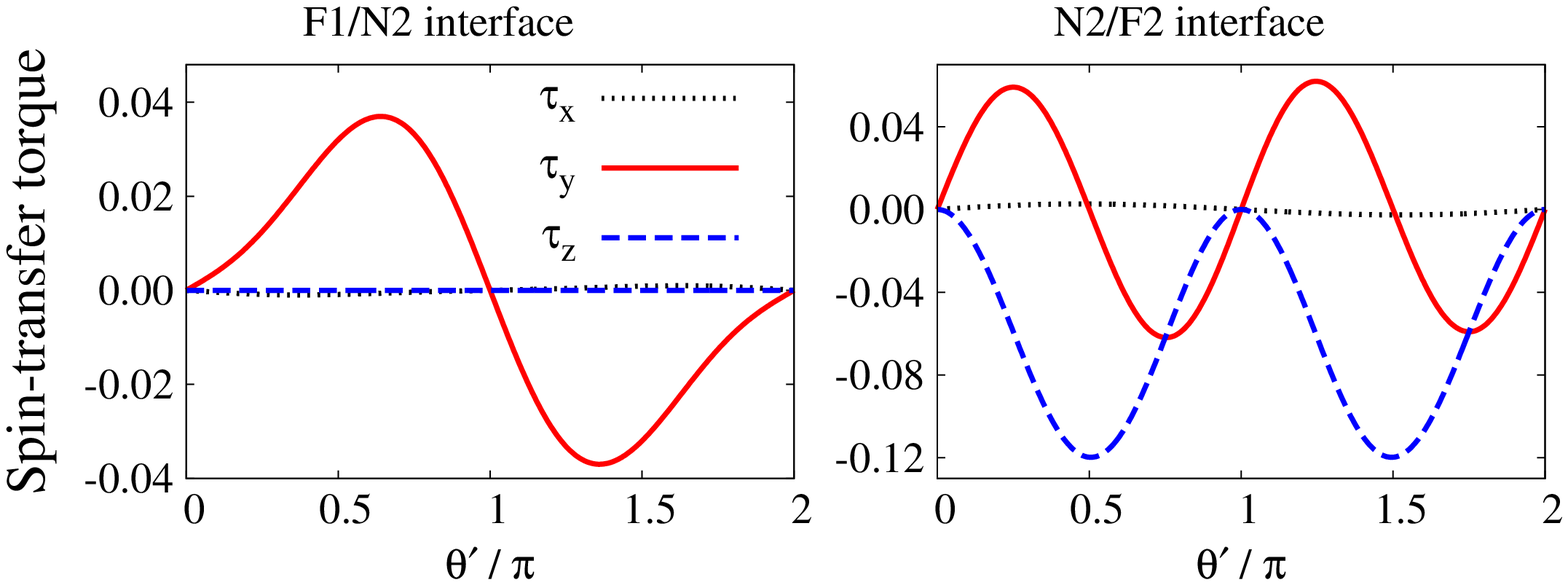}
    \caption{\label{Fig:stt2} (color online)
             Angular dependence of the cartesian STT
             components in the units of $\hbar I/ |\eq| $ acting on F1/N2 and N2/F1 interfaces when
             $\bhS_0 = \bhS_1 = \ez$ and $\bhS_2$ is rotated from $\ez$ in the layer's plane by an
             angle $\theta'$.}
  \end{figure}

  As will be shown in the following, CFL is usually not switched as a rigid
  structure,
  but generally forms a configuration which deviates from the antiparallel one.
  Figure~\ref{Fig:stt2} shows how the STT components at the $\F{1}$/$\N{2}$ and $\N{2}$/$\F{2}$ interfaces
  vary when
  $\bhS_2$ is rotated from $\ez$ by an angle $\theta'$, while $\bhS_1$
  remains
  fixed and is parallel to $\bhS_0 = \ez$.
  In such a case, the torque acting at $\N{1}$/$\F{1}$ interface remains zero,
  as $\bhS_1$ stays collinear to $\bhS_0$.
As before, the out-of-plane components are also negligible in
comparison to the in-plane ones.
  The in-plane components of STT reveal standard angular dependence at both interfaces.
  The amplitude of STT at the internal interfaces of CFL is comparable to that
  acting at the $\N{1}$/$\F{1}$ interface in the case of noncollinear configuration of $\bhS_0$ and
  $\bhS_1$, when $\bhS_2$ is fixed in the direction antiparallel to
  $\bhS_0$.

  When $\bhS_1$ is noncollinear to $\bhS_0$, the spin accumulation in $\N{1}$ layer
  increases and consequently the amplitude of STT at $\F{1}$/$\N{2}$ and $\N{2}$/$\F{2}$ decreases.
  In turn, when $\bhS_1$ is antiparallel to $\bhS_0$, the STT inside the CFL structure
  is reduced by more than a factor of 2.
  Nonetheless, the STT acting at the internal interfaces of the studied CFL layers
  might have a significant effect on their current-induced dynamics and
  switching process, provided the magnetic configuration of CFL
  might deviate remarkably from its initial antiparallel configuration.

\subsection{Synthetic antiferromagnet}

  First, we examine dynamics of the SyAF free layer.
  From symmetry we have $\hrkky{1} = \hrkky{2}\equiv  H_{\rm RKKY} $, and we
  have set $H_{\rm RKKY} =2\,\kOe$, which corresponds to
  $\Jrkky \sim -0.6\,\mathrm{mJ}/\mathrm{m}^2$.
  We have performed a number of independent numerical simulations modelling
  SyAF dynamics induced by constant  current and constant in-plane
  external magnetic field. The latter is assumed to be smaller than the critical field for
  transition to spin-flop phase of SyAF. Accordingly, each simulation started from an initial state
  close to $\bhS_1 = -\bhS_2 = -\ez$.
  To have a non-zero initial STT for $\bhS_1$, both spins of the SyAF  have been
  tilted by $1^{\circ}$ in the layer plane so that they remained collinear.

\begin{figure}[t]
    \centering
    \includegraphics[width=8cm]{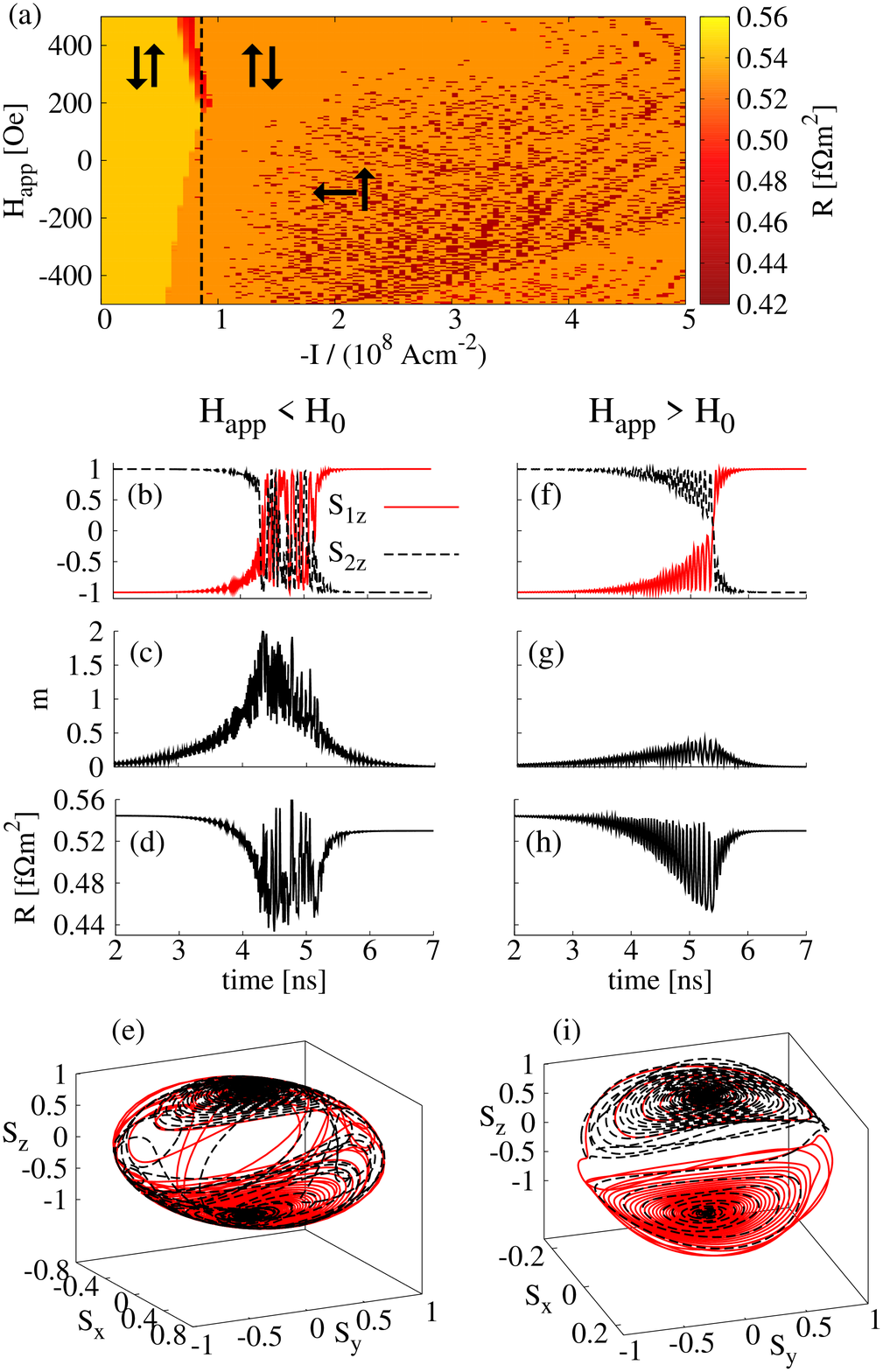}
    \caption{\label{Fig:SAF_dynam1} (color online)
             (a) Averaged resistance of
             Cu-IrMn(10)/Py(8)/Cu(8)/Co(2)/Cu(1)/Co(2)-Cu spin valve pillar
             with a SAF free layer as a function of current density and
             applied magnetic field.
             Examples of switching processes at
             $I = -1.0\times 10^{8} \,\mathrm{A cm}^{-2}$ and
             $\happ = -400\,\Oe$ (b -- e) and $\happ = 400\,\Oe$ (f -- i).
             Figures (b) and (f) show dynamics of $z$-components of both spin
             moments, (c) and (g) present the overal magnetization of the free layer,
             (d) and (h) show the corresponding variation of pillar resistance,
             (e) and (i) show the spin trajectories of $\bhS_1$ (red
             solid line) and $\bhS_2$ (black dashed line) in the time interval from $0$ to $10\,{\rm ns}$,
             where switching takes place.
             }
  \end{figure}

  From the results of numerical simulations we have constructed a map of time-averaged
  resistance, shown in Fig.~\ref{Fig:SAF_dynam1}(a).
  The resistance has been averaged in the time interval of $30\,\ns$
  following initial $50\,\ns$ equilibration time of the dynamics.
  The diagram shows only that part of the resistance, which depends on magnetic configuration,
  and   hence varies with CFL dynamics~\cite{Gmitra2009:PRB}.
  The constant part of resistance, due to bulk and interfacial
  resistances   of the studied structure, has been calculated to be as large as $\Rsp = 19.74\,\fOm$.
  For the assumed initial configuration,
  magnetic dynamics has been observed only for negative current density.
  When the current is small, no dynamics is
  observed since
  the spin motion is damped into the closest collinear state
  ($\bhS_1 = -\bhS_2 = -\ez$, marked as $\du$) of high resistance.
  After exceeding a certain threshold value of current density,
  there is a drop in the averaged resistance,
  which indicates  current induced dynamics of the SyAF free layer.
  Figures~\ref{Fig:SAF_dynam1}(b) and (f) show that this drop is associated
  with switching of the whole SyAF structure into an oposite state
  ($\bhS_1 = -\bhS_2 = \ez$, marked as $\ud$)

  From Fig.4(a) follows that  the threshold current for dynamics onset
  markedly depends on the applied field and
  reaches maximum at a certain value  of $\happ$, $\happ = \hsaf$.
  Furthermore, it appears that mechanisms of the switching process
  for $\happ < \hsaf$ and $\happ > \hsaf$ are qualitatively different.
  To distinguish these two mechanisms, we present in Figs.~\ref{Fig:SAF_dynam1}(b -- i)
  basic characteristics of switching,
  calculated for $I = -1.0\times 10^{8} \mathrm{Acm}^{-2}$ and for fields
  $\happ = -400\,\Oe$, which is below $\hsaf$ [Figs.~\ref{Fig:SAF_dynam1}(b -- e)],
  and $\happ = 400,\Oe$, which lies above $\hsaf$ [Figs.~\ref{Fig:SAF_dynam1}(f -- i)].
  Figs.~\ref{Fig:SAF_dynam1}(b) and (f) present time evolution of the z-components of both spins.
  To better understand the SyAF dynamics, in Figs.~\ref{Fig:SAF_dynam1}(c) and (g)
  we plotted the amplitude of overal SyAF magnetization, defined as $m = |\bhS_1 + \bhS_2|$.
  This parameter vanishes for antiparallel alignment of both spins of CFL, but becomes nonzero
  when the configuration deviates from the antiparallel one.
  Magnetization of SyAF is also a measure of the CFL coupling to external magnetic field.
  Futhermore, Figs.~\ref{Fig:SAF_dynam1}(d) and (h) show the corresponding time variation of the
  resistance, $R$, which might be directly extracted from experimental measurements as well.
  In addition, in Figs.~\ref{Fig:SAF_dynam1}(e) and (i) we show the trajectories
  of $\bhS_1$ and $\bhS_2$ in the real space taken from the time interval from
  $t = 0$ to $10\, {\rm ns}$. In addition,
  from Figs.~\ref{Fig:SAF_dynam1}(a) it has been found
  that the point where the threshold current reaches its maximum is located at $\hsaf \simeq \Hiz{0 2}$,
  which indicates its relation to magnetostatic interaction of F2 and fixed polarizer.
  This also has been confirmed by analogical simulations disregarding
  the magnetostatic coupling between magnetic layers, which resulted in similar diagram, but with
  $\hsaf = 0$ (not shown). This fact significantly facilitates understanding the mechanism of SyAF switching.

  The initial configuration assumed above was
  $-\bhS_1 = \bhS_2 \simeq \bhS_0$ with $\bhS_0 = \ez$ ($\downarrow\uparrow$).
  When the magnitude of current density is large enough and $I < 0$,
  orientation of $\bhS_1$ becomes unstable
  and $\bhS_1$ starts to precess with small angle around $-\ez$.
  Initial precession of $\bhS_1$ induces precession of $\bhS_2$ --  mainly via the RKKY coupling.
  Generally, response  to the exchange field is slower than
  current-induced dynamics.  Therefore, a difference in precession phase of $\bhS_2$ and $\bhS_1$
  appears, and configuration of SyAF
  deviates from the initial antiparallel one.
  This in turn enhances the STT acting on F2, which tends to switch $\bhS_2$.
  Its amplitude, however, is  small in comparison to the strong RKKY coupling.
  Further scenario of the dynamics depends then on the external magnetic field.
  When $\happ < \hsaf$ [Figs.~\ref{Fig:SAF_dynam1}(b -- e)]
  the Zeeman energy of $\bhS_2$ has a maximum in the initial state
  and external magnetic field tends to switch $\bhS_2$ to the opposite orientation.
  Competition between the torques acting on SyAF results in out-of-plane precessions of both spins.
  After several precessions $\bhS_1$ reaches the opposite static state, which is stable
  due to STT. In turn, $\bhS_2$ is only slightly affected by STT,
  and its dynamics is damped in the external magnetic and RKKY exchange fields.
  In contrast, when $\happ > \hsaf$ [Figs.~\ref{Fig:SAF_dynam1}(f -- i)], Zeeman energy of F2
  has a local minimum in the initial state, which stabilizes $\bhS_2$.
  Therefore, in a certain range of current density, SyAF does not switch but remains in self-sustained
  coherent in-plane precessions (red area in the upper part of Fig.~\ref{Fig:SAF_dynam1}(a)).
  For a sufficient current density, the SyAF structure becomes destabilized and
  the precessional angle increases until the spins pass the $(x,y)$-plane.
  Consequently, the precessional angle decreases and spins of the SyAF  are stabilized in the
  opposite state ($\uparrow\downarrow$).
  Moreover, as shown in Fig.~\ref{Fig:SAF_dynam1}(c), the switching process for $\happ < \hsaf$ is
  connected with high distortion of SyAF configuration, where $m$ in a certain point reaches its maximum
  value (corresponding to parallel orientation of both spins).
  Contrary, the $m$ remains small for $\happ > \hsaf$ [Figs.~\ref{Fig:SAF_dynam1}(g)],
  and the effective magnetic moment of the free layer
  stays smaller than magnetic moment of a single layer. This might
  play an important role in applications of spin-torque devices based on CFLs.

  The two  switching mechanisms described above dominate the current-induced dynamics when
  the current density is close to the dynamics threshold.
  For higher current densities, the nonlinearities in SyAF dynamics become more
  pronounced,
  which results in bistable behavior of the dynamics, especially for $\happ < \hsaf$ and
  $I \gtrsim 10^8 \mathrm{Acm}^{-2}$.
  In that region, the number of out-of-plane precessions before SyAF switching increases with the current density.
  However, their precessional angle increases in time and consequently
  $\bhS_1$ might reach an out-of-plane static point slightly tilted away from the $\ex$ direction
  while $\bhS_2 = \ez$ remains in the layers plane.
  The out-of-plane static states (marked as $\leftarrow\uparrow$)
  have small resistance and appear as dark red spots in the diagram shown in
  Fig.~\ref{Fig:SAF_dynam1}(a).

  In addition, from the analysis  of the dispersion of pillar resistance (not shown) one finds
  that except of a narrow region close to the dynamics threshold with
  persistent in-plane precessions, no significant steady-state dynamics of SyAF element appears.
  As will be shown below, such a behavior might be observed
  when CFL becomes asymmetric (SyF free layer).

\subsection{Synthetic ferrimagnet}

\begin{figure}
    \centering
    \includegraphics[width=8cm]{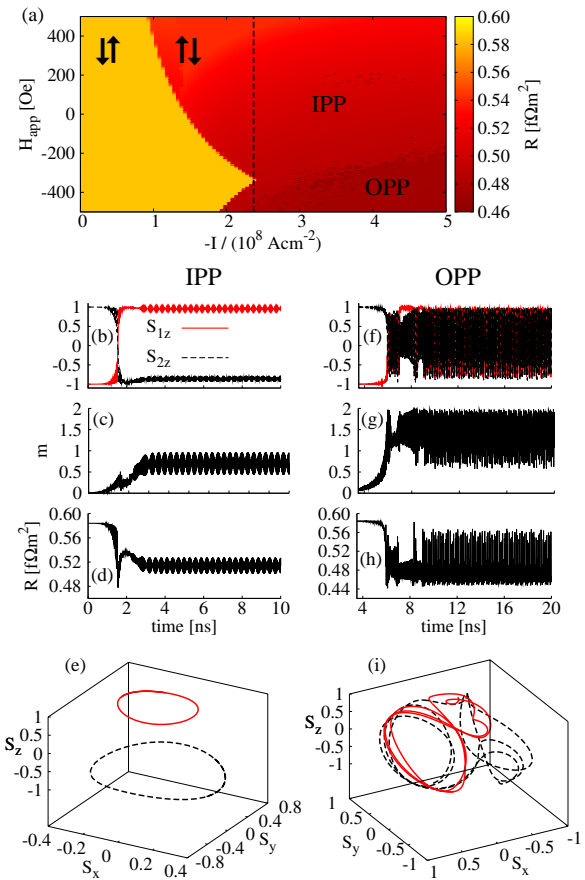}
    \caption{\label{Fig:SF_dynam1} (color online)
             (a) Averaged resistance of
             Cu-IrMn(10)/Py(8)/Cu(8)/Co(4)/Cu(1)/Co(2)-Cu spin valve pillar
             with a SyF free layer, presented as a function of current density and
             applied magnetic field.
             Examples of current-induced dynamics for $I = -3\times 10^8\,
             \mathrm{Am}^{-1}$ and $\happ = 200\,\Oe$ (b -- e) and
             $\happ = -400\,\Oe$ (f -- i).
             Panels (b) and (f) show dynamics of $z$-components of $\bhS_1$ and $\bhS_2$,
             (c) and (g) present the overal magnetization of the free layer,
             (d) and (h) show the corresponding variation of pillar resistance,
             (e) and (i) show  spin trajectories of $\bhS_1$ (red
             solid line) and $\bhS_2$ (black dashed line)
             taken from a time interval as large as $30\, {\rm ns}$
             after $100\, {\rm ns}$ of initial equilibration.}
  \end{figure}

  Let us study now spin valve with SyF as a free layer, assuming  $d_1 = 4\,\nm$ and $d_2 = 2\,\nm$.
  Accordingly, $\hrkky{2}$ remains $2\,\kOe$ while $\hrkky{1}$ is reduced to $1\,\kOe$.
  As in the case of  SyAF, from the averaged time-dependent part of the pillar resistance we have constructed a diagram
  presenting current-induced dynamics,
  see Fig.~\ref{Fig:SF_dynam1}(a).
  The static part of resistance is now $\Rsp = 19.80\,\fOm$.
  The diagram has some features similar to those studied in the previous subsection.
  However, the maximum critical current is shifted towards negative values of
  $\happ$,
  even if magnetostatic interaction  between magnetic layers is neglected.
  This asymmetry is caused by the
  difference in exchange and demagnetization fields acting
  on  layers $\F{1}$ and $\F{2}$.
  Moreover, this difference leads to more complex dynamics of the CFL's spins than that in the case of SyAF.

  Generally, there are several dynamic regimes to be distinguished.
  The first one is the region of switching from $\du$ configuration to the opposite
  one, $\ud$,  which is located at largest values of $\happ$ in the diagram.
  Mechanism of the switching is similar to that of SyAF shown in
  Figs.~\ref{Fig:SAF_dynam1}(f -- i),
  where CFL changes its configuration just {\it via} in-plane precessional states
  with a small value of $m$ (weak distortion of the antiparallel alignment of $\bhS_1$ and $\bhS_2$).
  Furthermore, the darker area above $\hsf$ indicates one of the possible self-sustained dynamic regimes of
  SyF, i.e. the in-plane precessions (IPP); see Figs.~\ref{Fig:SF_dynam1}(b -- e).
  This precessional regime starts directly after the SyF switching, and $\bhS_1$ and $\bhS_2$ precess
  around $\ez$ and $-\ez$, respectively.
  Due to different effective fields in $\F{1}$ and $\F{2}$, and energy gains due to
  STT,
  the spins precess with different precessional angles [Fig.~\ref{Fig:SF_dynam1}(e)]
  and consequently different frequencies.
  Because of the strong interlayer coupling and spin transfer between
  the layers,
  amplitudes of their precessions are periodically modulated in time.
  This modulation appears also in the time dependence of pillar resistance.
  Conversely, below $\hsf$ the dynamic is dominated by large angle out-of-plane
  precessions (OPP) of both spins, as shown in Figs.~\ref{Fig:SF_dynam1}(f -- i).
  This dynamic state is connected with a strong distortion of the antiparallel CFL configuration, i.e. large value of $m$,
  and large variation of the resistance.
  From Fig.~\ref{Fig:SF_dynam1}(i) one can see that trajectories of $\bhS_1$ and $\bhS_2$ are rather
  complicated including both IPP and OPP regimes with dominant OPP.

\subsection{Power spectral density}
\label{SSec:PSD}

  From the analysis of current-induced dynamics we found that
  self-sustained dynamics in structures with SyF free layer
 is much richer than that in systems with SyAF
  free layer [see Figs.~\ref{Fig:SF_dynam1}(b -- i)].
  Therefore, in this section we restrict ourselves  to
  dynamic regimes of the SyF free layer only.
  More specifically, we shall examine the power spectral density (PSD) as a function of
  current density and external magnetic field.

  In the simulation we started  from $I = 0$ and changed current density in steps
  $\Delta I = 10^{6}\,\mathrm{Acm}^{-2}$ at a fixed applied field.
  As before, to protect the SyF dynamics from collapsing into collinear static state, we assumed
  small thermal fluctuations corresponding to $\Teff = 5\,\K$.
  At each step we simulated the dynamics of coupled CFL's spins
  and calculated PSD.
  As in Ref.~\onlinecite{Urazhdin2009:PRB}, we assumed that the input current is split between
  a load with resistance $\RL$ and pillar with resistance $\Rsp + R(t)$.
  Hence, voltage on the pillar has been calculated as $U(t) = I R(t) / [1 + \Rsp / (\RL S) ]$,
  where we assumed $\RL = 50\, \Omega$, and $S$ is
  the cross-section of the pillar (ellipsoid with the major and minor axes equal to $130$ and $60$~nm, respectively).
  Then, at a given $I$ we calculated voltage in the frequency domain, $U(f)$,
  using fast Fourier transformation over the period $t_{\rm FFT} = 50\,\ns$ following
  the equilibration time of $t_{\rm eq} = 30\, \ns$.
  The power spectral density has been defined as
  $\PSD(f)= 2 U^{2}(f) / (\RL\, \Delta f)$, where $\Delta f = 1 / t_{\rm FFT}$.

  \begin{figure}[t]
    \centering
    \includegraphics[width=.9\columnwidth]{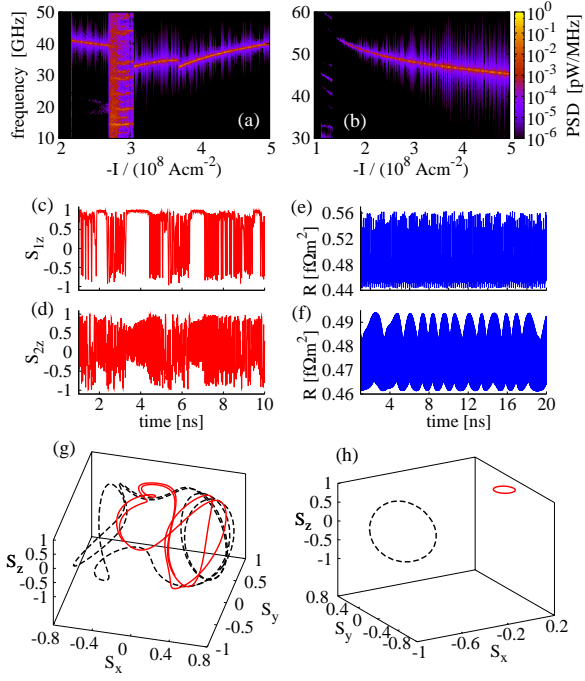}
    \caption{\label{Fig:psd}
             Power spectral density calculated for the spin valve with SyF free layer
             at $\Teff = 5\,\K$ and $\happ = -400\,\Oe$ (a) and $200\,\Oe$ (b).
             (c) and (d) show steady time evolution of spins $z$-components in
             a time window of $10\,\ns$ after the equilibration at
             $\happ = -400\,\Oe$ and $I = -2.8 \times 10^{8}\, \mathrm{Acm}^{-2}$.
             (e) and (f) show steady time evolution of the time-dependent part of spin valve resistance
             in a time window of $20\,\ns$ after the equilibration at
             $\happ = -400\,\Oe$ and $I = -3.6 \times 10^{8}\, \mathrm{Acm}^{-2}$
             (both $\bhS_1$ and $\bhS_2$ precess out of layer plane) and
             $I = -3.8 \times 10^{8}\, \mathrm{Acm}^{-2}$
             ($\bhS_2$ performs out-of-plane precessions while $\bhS_1$ precesses in the layer's plane), respectively.
             Panels (g) and (h) depict trajectories of $\bhS_1$ (red solid line)
             and $\bhS_2$ (black dashed line) corresponding to
             resistance oscillations (e) and (f), respectivelly.}
  \end{figure}
  Figures~\ref{Fig:psd}(a) and (b) show PSD calculated
  at $\happ = -400\,\Oe$ and $200\,\Oe$, respectively. The former case corresponds to
  that part of the diagram in Fig.~\ref{Fig:SF_dynam1}(a), which includes OPP
  modes,
  while in the latter case we observed IPP only.
  Let us analyze first the situation in Fig.~\ref{Fig:psd}(a).
  When current passes through the corresponding threshold value, both spins start precessing in the layers' plane,
  similarly as shown in Fig.~\ref{Fig:SF_dynam1}(b).
  Apart from the main peak in the PSD at $f \simeq 40\,\GHz$,
  two additional minor peaks close to $f \simeq 20\,\GHz$ are visible.
  We attribute them to the oscillations of precessional amplitudes of both spins.
  With increasing amplitude of the current density, the precessional angles of both
  spins increase and their precessional frequencies slightly decrease.
  Moreover, with increasing current the frequencies of the minor peaks become
  closer and closer, until they finally coincide.
  At this point the PSD becomes widely distributed along the whole range of observed frequencies,
  which is an evidence of noisy variation of the resistance.
  An example of spin dynamics in this region is shown in Figs.~\ref{Fig:psd}(c) and (d)
  which have been taken in a time window as large as $10\,\ns$ after the equilibration period
  for $\happ = -400\,\Oe$ and $I = -2.8\times 10^{8}\,\mathrm{Acm}^{-2}$
  [within the broad feature of PSD in Fig.~\ref{Fig:psd}(a)].
  Firstly, the figures show that $\bhS_2$ starts to perform
  out-of-plane precessions
  as a result of the competition between  STT and RKKY coupling.
  Secondly, one can note thermally activated random transitions of $\bhS_1$
  between OPP and IPP modes.
  These random transitions modify OPP precessions of $\bhS_2$ as well.
  Simultaneous dynamics of both spins causes chaotic variation
  of spin valve resistance and broadens the PSD.
  The quasi-chaotic feature of the spin dynamics in this range of current densities
  can be seen also on the spin trajectories, which cover almost the whole sphere (not shown).

  Further increase in current density leads to stabilization of the OPP mode of $\bhS_1$.
  Hence spin valve resistance becomes more periodic [see Fig.~\ref{Fig:psd}(e)]
  and PSD reveals a narrow peak again.
  Since both spins perform rather complicated dynamics including IPP but dominated by
  OPP regime [see Fig.~\ref{Fig:psd}(g)], we observe a blue-shift in PSD with current,
  which is connected with a decrease in the precessional angles.
  However, at a certain value of $I$ we notice an abrupt drop in the peak's frequency.
  At this current density the STT acting on the left
  interface of layer $\F{1}$  starts to dominate the dynamics of $\bhS_1$ and
  enables only small angle IPPs along the $\bhS_0$ direction, which
  modifies the trajectory of $\bhS_2$.
  $\bhS_2$ still remains in the OPP regime [see Fig.~\ref{Fig:psd}(h)] and hence the blue-shift with current appears.
  The fact that IPP of $\bhS_1$ still influence the dynamics of the whole SyF
  is also shown in Fig.~\ref{Fig:psd}(f), which presents the dynamic part of the spin valve
  resistance at $I = -3.8 \times 10^8 \mathrm{Acm}^{-2}$ and $\happ = -400\,\Oe$.
  As a result of IPPs of $\bhS_1$, amplitude of the resistance varies periodically.
  In addition, comparison of Figs.~\ref{Fig:psd}(e) and (f) shows that
  the simultaneous OPPs of both spins lead to stronger variation of the resistance
  than in the case when the layers are in the IPP state.

  Contrary, at $\happ = 200\, \Oe$ one observes only IPP modes of both spins
  similar to those shown in Fig.~\ref{Fig:SF_dynam1}(e).
  The in-plane precessional angle increases with current density and hence
  the peak frequency in PSD decreases and becomes broader.
  In real systems, however, one might expect the peaks narrower than
  those obtained in the  macrospin simulations,
  as observed in standard spin valves with
  a simple free layer~\cite{Sankey2005:PRB,JVKim2006:PRB}.


\section{Critical currents}
\label{Sec:Currents}

  First, we derive some approximate expressions for critical current density
  needed to induce dynamics of CFL, derived from linearized  LLG equation.
In metallic structures, the out-of-plane torque components are
generally  much smaller than the in-plane
  ones,
  and therefore will be omitted in the analytical considerations of this section
  ($b_1^{(0)}, b_1^{(2)}, b_2^{(1)} \rightarrow 0$).

  The coupled LLG equations in spherical coordinates can be then written as
  \begin{equation}
  \label{Eq:LLG_spherical}
    \der{}{\tilde{\bf S}}{t} = \frac{1}{1 + \alpha^2}\, \bar{\bf M} \cdot \tilde{\bf v}\,,
  \end{equation}
  where $\tilde{\bf S} = (\theta_1, \phi_1, \theta_2, \phi_2)^{\rm T}$ is a 4-dimensional column vector
which describes spin orientation in both layers constituting the
CFL, and
  $\tilde{\bf v} = (v_{1 \theta}, v_{1 \phi}, v_{2 \theta}, v_{2 \phi})^{\rm T}$ stands
  for the torque components, $v_{i \theta} = {\bm \Gamma}_{i} \cdot \et{i}$ and
  $v_{i \phi} = {\bm \Gamma}_{i} \cdot \ep{i}$, with
  $\ep{i} = (\ez \times \bhS_i) / \sin\theta_i$ and
  $\et{i} = (\bhS_i \times \ep{i}) / \sin\theta_i$
  denoting unit vectors in local spherical coordinates associated with $\bhS_i$.
  In turn, the  $4 \times 4$ matrix $\bar{\bf M}$ takes the form
  \begin{equation}
    \bar{\bf M} =
    \begin{pmatrix}
      1 & \alpha & 0 & 0 \\
      -\alpha / \sin\theta_1 & 1 / \sin\theta_1 & 0 & 0 \\
      0 & 0 & 1 & \alpha \\
      0 & 0 & -\alpha / \sin\theta_2 & 1 / \sin\theta_2
    \end{pmatrix}.
  \end{equation}
  Static points of the CFL dynamics have to satisfy $v_{i \theta} = 0$ and
  $v_{i \phi} = 0$ for both $i = 1$ and $i = 2$.
  These conditions are obeyed in all collinear configurations, i.e.
  $\theta_i = 0, \pi$.
  Additional four trivial static points can be found in the out-of-plane configurations with
  $\theta_i = \pi/2$ and $\phi_i = 0, \pi$.

  Following Ref.~\onlinecite{Bazaliy2004:PRB} we linearize Eq.~(\ref{Eq:LLG_spherical}) by
  expanding  $\tilde{\bf v}$ into a series around the static points, which leads to
  \begin{equation}
    \der{}{\tilde{\bf S}}{t} = \frac{1}{1 + \alpha^2}\, \bar{\bf M} \cdot \bar{\bf J}
                               \cdot \tilde{{\bm \delta}\!{\bf
                               v}}\, ,
  \end{equation}
  where $\bar{\bf J}$ is a Jacobian matrix of
  $\partial \tilde{v}_i / \partial \tilde{S}_j$ components.
The matrix product $\bar{\bf M} \cdot \bar{\bf J}$  defines here
the dynamic matrix
  $\bar{\bf D} = \bar{\bf M} \cdot \bar{\bf J}$.
  This matrix allows one to study stability of
  CFL's spins in their static points.
  If $\Tr{\bar{\bf D}}$ is negative, the static point is stable,
  otherwise it is unstable.
  Hence, the condition for critical current is~\cite{Wiggins1990:Springer}
  $\Tr{\bar{\bf D}} = 0$.

  To obtain threshold current for dynamics onset of individual spins in the CFL,
  we assume first that one of the spins is fixed in its initial position and investigate
  stability of the second spin.
  The dynamic matrix $\bar{\bf D}$ becomes then reduced to a $2 \times 2$ matrix.
  Considering initial position of SAF with $\bhS_1 = -\bhS_2 = -\ez$
  (i.e. $\theta_1 = \pi$ and $\theta_2 = 0$), marked as $\du$, and polarizer $\bhS_0 =
  \ez$,
  the stability condition leads to the following critical currents $\Icdu{1}$ and $\Icdu{2}$
  for $\bhS_1$ and $\bhS_2$, respectively:
  \begin{equation}
    \Icdu{1} = -\alpha \frac{\mu_0 \Ms d_1}{a_{1}^{(0)} + a_{1}^{(2)}}
    \left[-\hext^{1 \du} + \hani -\hrkky{1} + \Hd{1} \right]\, ,
  \label{Eq:CritCurr1_du}
  \end{equation}
  with $\hext^{1 \du} = \happ - \Hiz{01} - \Hiz{21}$, and
  \begin{equation}
    \Icdu{2} = -\alpha \frac{\mu_0 \Ms d_2}{a_{2}^{(1)}}
    \left[\hext^{2 \du} + \hani -\hrkky{2} + \Hd{2} \right]\, ,
  \label{Eq:CritCurr2_du}
  \end{equation}
  with $\hext^{2 \du} = \happ - \Hiz{02} - \Hiz{12}$.
  In both above expressions $a_1^{(0)}$, $a_1^{(2)}$, and $a_2^{(1)}$ are
  taken in the considered static point, while
  the demagnetization field for the $i$-th layer is given by
  \begin{equation}
    \Hd{i} = \frac{\Hdx{i} + \Hdy{i}}{2} - \Hdz{i}\,.
  \end{equation}
  Analogically, one can derive similar formulas for critical currents
  in the opposite $(\ud)$ magnetic configuration of the CFL.

  Now we relax the assumption that one of the spins is fixed, and consider both spins of the CFL as free.
  Then, we calculate the trace of the whole $4 \times 4$ matrix, which leads to
  the following
  expression for critical current destabilizing  the whole CFL:
  \begin{equation}
  \label{Eq:IcCFL}
    \begin{split}
      \Icdu{\rm CFL} &=
      -\alpha \frac{\mu_0 \Ms d_1 d_2}{d_2 (a_1^{(0)} + a_1^{(2)}) + d_1 a_2^{(1)}}\,
      \biggl[ \hext^{\du} + 2 \hani \\
      &\quad - \hrkky{1} - \hrkky{2} + \Hd{1} + \Hd{2} \biggr]\,,
    \end{split}
  \end{equation}
  where  $\hext^{\du} = H_{z}^{01} - H_{z}^{02} + H_{z}^{21} + H_{z}^{12}$.
  Since the spins of CFL are antiparallel in  the considered static point,
  $\Icdu{\rm CFL}$ is independent of external magnetic field.
  The above equation describes the critical current at which the CFL
  is destabilized  as a rigid structure (unaffected by external magnetic field along the $z$-axis).

  Numerical calculations presented below show that critical current is usually smaller than that given by
  Eq.(12).
  Apparently, as shown by numerical simulations,
  there is a  phase shift in initial precessions of $\bhS_1$ and $\bhS_2$.
  Such a phase shift slightly perturbs initial antiparallel configuration and
  might reduce the critical current for the dynamics onset.

  Similar formula also holds for the opposite configuration ($\ud$), where
  the critical current is given by
  \begin{equation}
  \label{Eq:IcCFL_opp}
    \begin{split}
      \Icud{\rm CFL} &=
      \alpha \frac{\mu_0 \Ms d_1 d_2}{d_2 (a_1^{(0)} - a_1^{(2)}) - d_1 a_2^{(1)}}\,
      \biggl[ -\hext^{\ud} + 2 \hani \\
      &\quad - \hrkky{1} - \hrkky{2} + \Hd{1} + \Hd{2} \biggr]\,,
    \end{split}
  \end{equation}
  with  $\hext^{\ud} = H_{z}^{01} - H_{z}^{02} - H_{z}^{21} - H_{z}^{12}$.

   Now we discuss the theoretical results on critical currents in the context of
  those following from numerical simulations.
  Let us consider first the critical currents for individual spins
  of the SyAF free layer, assuming that
  the second spin remains stable in its initial position, Eqs.~(\ref{Eq:CritCurr1_du})
  and~(\ref{Eq:CritCurr2_du}).
  The corresponding results obtained from the formula derived above are presented in Table~\ref{Tab:crit},
  where we have omitted a weak dependence on $\happ$.
  For the studied structure with  SyAF free layer,
  $\Icdu{1}$ is negative while $\Icdu{2}$ is positive.
  From our analysis follows, that the current density at which dynamics appears in the simulations (Fig.~\ref{Fig:SAF_dynam1}(a))
  is higher than that given by $\Icdu{1}$.
  However, we checked numerically that the critical value $\Icdu{1}$ agrees with the critical current
  obtained from simulations when assuming
  $\bhS_1$ as free and  fixing $\bhS_2$ along $\ez$.

  \begin{table}[h]
    \begin{ruledtabular}
    \begin{tabular}{lcccc}
       & \multicolumn{2}{c}{SyAF} & \multicolumn{2}{c}{SyF}\\
       & $\du$ & $\ud$ & $\du$ & $\ud$ \\ \hline
      $I_{{\rm c} 1}$ & -0.31  & 0.43 & -0.54 & 0.63\\
      $I_{{\rm c} 2}$~ & 0.98 & -0.46 & 2.33 & -0.78\\
      $I_{{\rm c} CFL}$~ & -0.86 & 0.87 & -2.37 & -9.18
      \end{tabular}
    \end{ruledtabular}
    \caption{\label{Tab:crit} Critical current densities in the units of
             $10^8\,\mathrm{Acm}^{-2}$, calculated according to
             equations~(\ref{Eq:CritCurr1_du}), (\ref{Eq:CritCurr2_du}),
             and~(\ref{Eq:IcCFL}) for both SyAF and SyF free layers.}
  \end{table}

  Following the above discussion of the CFL dynamics,
  one can understand the shift of the threshold current as follows.
  Initially, when the current density exceeds $\Icdu{1}$, $\bhS_1$ becomes destabilized.
  Then, $\bhS_2$ responses to the initial dynamics of $\bhS_1$ with similar coherent precession.
  However, $\bhS_2$ should still be stable in its initial position at this current density and
  common precessions of the two coupled spin moments damps the initial dynamics. Accordingly,
  SyAF ends up in the closest static state ($\du$).
  However, as the current density increases, the initial precessions of $\bhS_1$ become more pronounced,
  which in turn
  means that the initial antiparallel configuration becomes distorted and
  $\bhS_2$ becomes destabilized.
  This results in coupled dynamics of both spins and finally leads to switching of the SyAF structure.

  On the other hand, we have also calculated the critical current for the whole SyAF structure according to
  Eq.~(\ref{Eq:IcCFL}), and for the given structure we got
  $\Icdu{CFL}$
  shown in Fig.~\ref{Fig:SAF_dynam1}(a) by the dashed vertical line (see also Table I).
  Equation~(\ref{Eq:IcCFL}) describes stability of the whole CFL, and since the
  interlayer coupling is strong, $\Icdu{CFL}$ corresponds to the current density
  at which both spins become destabilized
  simultaneously preserving their antiparallel orientation.
  As can be seen in Fig.~\ref{Fig:SAF_dynam1}(a), this is the
  maximum threshold current density  for current-induced dynamics.
  Because the rigid structure consisting of two antiparallel spins is not influenced by
  an external homogeneous magnetic field, there is no dependence of $\Icdu{CFL}$ on $\happ$.
  Nevertheless, from our numerical simulation follows that the threshold current
  for the SyAF dynamics, $\Itr$, obeys the condition
  $|\Icdu{1}| < |\Itr| < |\Icdu{CFL}|$, provided that $|\Icdu{1}| < |\Icdu{2}|$
  or (as in our case) $\Icdu{2}$ has different sign.

\begin{figure}[t]
    \centering
    \includegraphics[width=.9\columnwidth]{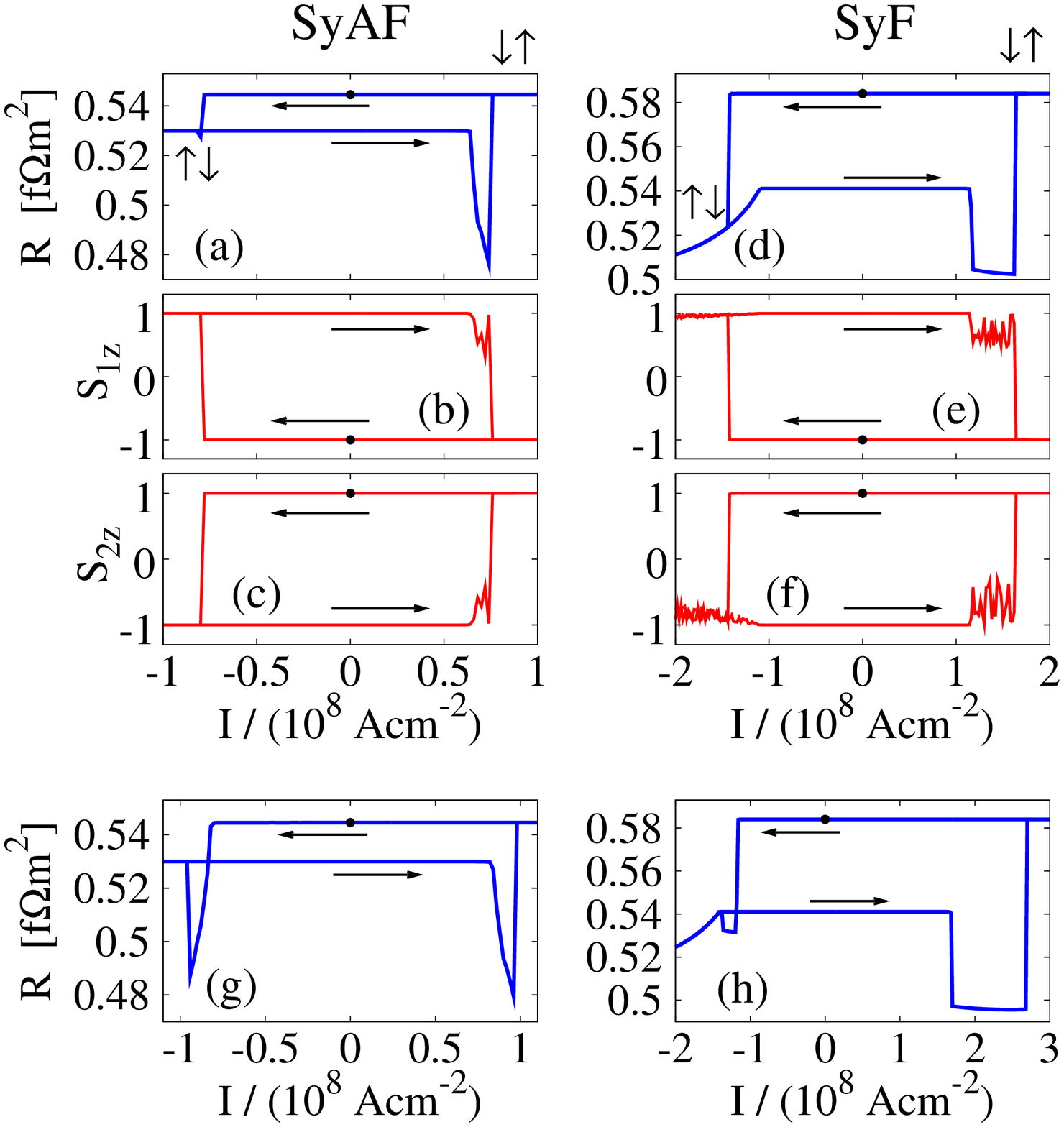}
    \caption{\label{Fig:hyster}
             Hysteresis loops of the resistance for the studied pillars with SyAF (a)
             and SyF (d) free layers.
             Panel (b) and (c) depict spin dynamics of $\bhS_1$ and $\bhS_2$ in SyAF, respectivelly,
             corresponding to resistance loop (a).
             Panels (e) and (f) show dynamics of $\bhS_1$ and $\bhS_2$ in SyF, respectivelly,
             corresponding to resistance loop (d).
             The initial point of each hysteresis loop is marked with a dot.
             The arrows indicate direction of the current change.
             Figures (g) and (h) correspond to the upper parts
             of (a) and (d), in which however the effects due to magnetostatic field
             of the reference layer to the CFL spins have been omitted.}
  \end{figure}

  When the SyAF is in the $\ud$ configuration, the spin accumulation and spin current are
  different from those in the  $\du$ configuration (at the same
  voltage). This in turn leads to different spin torques, which
  is the reason why the critical currents destabilizing $\ud$ state are different from those for $\du$,
  as shown in Table~\ref{Tab:crit}.
  From the critical currents  one can expect
  relatively symmetric hysteresis with applied current in structures with SyAF.
  In contrast, $\Icud{CFL}$ for the SyF is negative, similarly as $\Icdu{CFL}$, but it is significantly
  larger,   which indicates lack of hysteresis.
  To compare switching of the SyAF and SyF free layers from the $\du$ to $\ud$ configurations  with the
  opposite one ($\ud$ to $\du$), we have simulated dynamics of the corresponding
  CFLs  assuming $\happ = 0$ and varying current density.
  The simulations have been performed in the quasistatic regime, i.e., for each value of current density
  the spin dynamics was first equilibrated for $50\,\ns$ and then
  averaged values of spin components and pillar resistance were calculated from the data taken for the next
  $30\,\ns$ of dynamics.
  In order to prevent the system from  collapsing into a static state with zero
  torque, we have included a thermal stochastic field corresponding to
  $\Teff = 5\,\K$ [see Eq.~(\ref{Eq:Dth})].
  Starting from $I = 0$ and going first towards negative currents we have constructed
  the current dependence of the
  averaged resistance and related $z$-components of both spins, as  shown in
  Fig.~\ref{Fig:hyster}.
  For both SyAF [Figs.~\ref{Fig:hyster}(a --c)] and SyF [Figs.~\ref{Fig:hyster}(d -- f)]
  free layers, one can see relatively symmetric hysteresis with the current density.
  In both cases direct switching from $\du$ to $\ud$ state  occurs at a current density
  comparable to $\Icdu{CFL}$.
  In contrast, in the case of SyF free layer, the second transition ($\ud$ to $\du$) appears at a current density
  which is very different from that predicted by the linearized LLG model.
  Moreover, in both cases switching from $\ud$ to $\du$ state does not appear
  directly, but through some precessional states.
  More precisely, as the positive current density increases, both spins start
  precessing
  in the layers' plane prior to switching.
  The in-plane precessions are connected with a significant drop in the resistance
  and with a reduction of the $s_z$-components.
  The range of IPP regime is particularly large in the case of SyF.
  From the analysis of spins' trajectories one may conclude
  that the angle of IPPs increases with increasing current
  density,
  and after exceeding a certain threshold angle CFL switches to the $\du$ configuration.

  The other factor giving rise to the the difference in
  switching from $\ud$ to $\du$ and  from  $\du$ to $\ud$ follows from the fact that  the
  magnetostatic interaction of the CFL's layers with the polarizer is
  different in the $\du$ and $\ud$ states.
  To prove this we have constructed analogical hysteresis loops for SyAF and SyF free layers
  disregarding magnetostatic interaction with the $\F{0}$ layer;
  see Figs.~\ref{Fig:hyster}(g) and (h).
  For both SyAF and SyF free layers we observe now large decrease in $R$
  for
  both switchings. This implies that both switchings are realized {\it via} in-plane
  precessions,
  in contrast to the case when $\F{0}$ influences the CFL dynamics via the corresponding magnetostatic field.
  While the hysteresis loop for SyAF remains symmetric, the  one for SyF becomes highly asymmetric.
  The asymmetry of SyF loop is due to a significant asymmetry of STT in $\ud$ and $\du$ states,
  which was previously shaded by the magnetostatic coupling with the layer $\F{0}$.


\section{Discussion and conclusions}
\label{Sec:Conclusion}

  We have studied current-induced dynamics of  SyAF and SyF composite free layers.
  By means of numerical simulations we identified variety of dynamical regimes.
  The most significant difference between dynamics of SyAF and SyF free layers concerns the evidence of
  self-sustained dynamics of both CFL spins.
  While in the case of SyAF only coupled in-plane precessions in a narrow window of external parameters
  ($\happ$ and $I$) are observed,
  SyF free layer reveals more complex and richer dynamics,
  with the possibility of coupled out-of-plane precessions
  which might be interesting from the application point of view.
  Furthermore, as shown by numerical simulations, both SyAF and SyF  are
  switchable
  back and forth without the need of external magnetic field.
  For SyAF element two possible ways of switching have been identified.
  Since they lead to different switching times, their identification might be crucial
  for optimization of switching in real devices with SyAF free layers.
  However, one has to note that the diagrams shown in Figs.~\ref{Fig:SAF_dynam1} and~\ref{Fig:SF_dynam1}
  may be changed when magnetization in CFL becomes non-homogeneous.

  A disadvantage of the studied structures is their relatively low efficiency of switching,
  i.e. high amplitude of critical current and long switching time.
  In order to show more sophisticated ways of tuning the CFL
  devices,
  we have analyzed critical currents derived
  from the linearized LLG equation.
  The formula~(\ref{Eq:IcCFL}) has been identified as the maximum value
  of critical current at which dynamics of the CFL structure should be observed.
  This formula reveals some basic dependence of critical current on spin valve
  parameters,
  and therefore might be useful as an initial tool for its tuning.
  However, in some cases non-linear effects in CFL dynamics might
  completely change the process of CFL switching,
  as shown by the presented numerical simulations.
  But the effects of non-linear dynamics go beyond the simple
  approach of linearized LLG equation, and
  their study requires more sophisticated nontrivial methods
  and/or numerical simulations.


\section*{Acknowledgements}
This work was supported by Polish Ministry of Science and Higher
Education as a research project in years 2010-2011, and partly by
EU through the Marie Curie
  Training Network SPINSWITCH (MRTN-CT-2006-035327).
  The authors thank M. Gmitra for helpful discussions.
  One of us (PB) also thanks L. L\'opez D\'{\i}az, E. Jaromirska, U. Ebels, and D. Gusakova
  for valuable suggestions.


\appendix

\section{Transformations of spin current}
\label{Sec:Transformations}

  Torque acting on the left interface of $\F{1}$ is calculated from $x$ and $y$ components of
  ${\bf j}'_{1} = \bar{\bf T}(\theta_1, \phi_1).{\bf j}_{1}$, where ${\bf j}_{1}$ is
  spin current vector in $\N{1}$ layer (written in global frame; shown in Fig.~\ref{Fig:Scheme}), and
  $\bar{\bf T}(\theta_1, \phi_1) = \bar{\bf R}_x (-\theta_1) \bar{\bf R}_z (\phi_1 - \pi/2)$, where
  $\bar{\bf R}_q (\alpha)$ is the matrix of rotation by angle $\alpha$ along the axis $q$ in the
  counterclockwise direction when looking towards origin of the coordinate system. 
  Hence ${\bf j}'_{1}$ components can be written as
  \begin{subequations}
    \begin{align}
      j'_{1x} &= j_{1x} \sin\phi_1 - j_{1y} \cos\phi_1\,,\\
      j'_{1y} &= (j_{1x} \cos\phi_1 + j_{1y} \sin\phi_1) \cos\theta_1 - j_{1z} \sin\theta_1\,,\\
      j'_{1z} &= (j_{1x} \cos\phi_1 + j_{1y} \sin\phi_1) \sin\theta_1 + j_{1z} \cos\theta_1\,,
    \end{align}
  \end{subequations}
  where $(\theta_1, \phi_1)$ are spherical coordinates of $\bhS_1$ in the global frame.
  Similarly, we define torques' amplitudes on the left interface of $\F{2}$ from the components of transformed
  spin current vector ${\bf j}'''_{2} = \bar{\bf T}(\theta_2, \phi_2).{\bf j}_{2}$.
  In this case, however, ${\bf j}_{2}$ is not written in the global frame, but in the local
  coordination system coordinate system connected with $\bhS_1$.
  To rotate local coordinate system of $\bhS_1$ to local coordinate system of $\bhS_2$ we need to
  know spherical angles $\theta_2$ and $\phi_2$ of vector $\bhS_2$ in the local coordinate system of $\bhS_1$.
  This might be done by transforming first $\bhS_2$ vector
  to local coordinate system of $\bhS_1$ as $\bhS'_2 = \bar{\bf T}(\theta_1, \phi_1) \cdot \bhS_2$
  and calculate its angles $\theta_2$ and $\phi_2$.
  Then we can calculate components of ${\bf j}'''_{2}$ similarly as for the left interface
   \begin{subequations}
    \begin{align}
      j'''_{2x} &= j_{2x} \sin\phi_2 - j_{2y} \cos\phi_2\,,\\
      j'''_{2y} &= (j_{2x} \cos\phi_2 + j_{2y} \sin\phi_2) \cos\theta_2 - j_{2z} \sin\theta_2\,,\\
      j'''_{2z} &= (j_{2x} \cos\phi_2 + j_{2y} \sin\phi_2) \sin\theta_2 + j_{2z} \cos\theta_2\,,
    \end{align}
  \end{subequations}

  Equation in $\N{2}$, which is adjacent non-magnetic interface from the right-hand side of $\F{1}$,
  are written in local coordinate system of $\bhS_1$.
  To apply the definition of $a_{12}$ and $b_{12}$ we need to rotate the local coordinate system
  so, that its $y$-axis will lie in the layer given by vectors $\bhS_1$ and $\bhS_2$.
  This might be done by single rotation of local coordinate system around its $z$-axis by angle
  $\phi_2 - \pi/2$, ${\bf j}''_{2} = \bar{\bf R}_z (\phi_2 - \pi/2).{\bf j}_2$, where
  \begin{subequations}
    \begin{align}
      j''_{2x} &= j_{2x} \sin\phi_2 - j_{2y} \cos\phi_2\,,\\
      j''_{2y} &= j_{2x} \cos\phi_2 + j_{2y} \sin\phi_2\,,\\
      j''_{2z} &= j_{2z}.
    \end{align}
  \end{subequations}
  Note, angle $\phi_2$ is calculated for vector $\bhS_2$ transformed into coordinate system of $\bhS_1$
  as in previous case.


\end{document}